\documentclass[12pt]{article}
\usepackage{jhep-mod}
\usepackage{bm}
\usepackage{amssymb,amsmath,amsthm}
\usepackage{mathrsfs}
\usepackage[utf8]{inputenc}
\usepackage{enumerate}
\hypersetup{colorlinks=true} 
\usepackage{appendix}
\usepackage{graphicx}
\usepackage{dcolumn}
\usepackage{bm}
\usepackage{multirow}
\usepackage{float}
\usepackage[normalem]{ulem}
\definecolor{purple}{rgb}{1,0,1}

\begin{document}

\title{\huge{Post-Newtonian particle physics in curved spacetime}}

\author{\Large Matt Visser}
\affiliation{School of Mathematics and Statistics, Victoria University of Wellington, \\
PO Box 600, Wellington 6140, New Zealand}
\emailAdd{matt.visser@sms.vuw.ac.nz}

\abstract{

\parindent0pt
\parskip7pt

In three very recent papers, (an initial paper by Morishima and Futamase, and two subsequent papers by  Morishima, Futamase, and Shimizu),  it has been argued that the observed experimental anomaly in the anomalous magnetic moment of the muon might be explained using general relativity. It is my melancholy duty to report that these articles are fundamentally flawed in that they fail to correctly implement the Einstein equivalence principle of general relativity. Insofar as one accepts the underlying logic behind these calculations (and so rejects general relativity) the claimed effect due to the Earth's gravity will be swamped by the effect due to Sun (by a factor of fifteen), and by the effect due to the Galaxy (by a factor of two thousand). In contrast, insofar as one accepts general relativity, then the claimed effect will be suppressed by an extra factor of [(size of laboratory)/(radius of Earth)]$^2$.  Either way, the claimed effect is not compatible with explaining the observed experimental anomaly in the anomalous magnetic moment of the muon.

\bigskip
{\sc Date:} 3 February 2018; \LaTeX-ed \today


}

\maketitle

\def\tr{{\mathrm{tr}}}
\def\diag{{\mathrm{diag}}}
\parindent0pt
\parskip7pt
\section{Introduction}

Morishima and Futamase~\cite{Morishima:2018a}, and  Morishima, Futamase, and Shimizu~\cite{Morishima:2018b,Morishima:2018c},  have argued that the observed experimental anomaly in the anomalous magnetic moment of the muon might be explained using curved-spacetime quantum physics. Unfortunately these articles do not correctly implement the Einstein equivalence principle, and so their methods and conclusions are in direct conflict with general relativity. The quick way to see that there is an issue is to note that the claimed effect depends on the absolute gravitational potential $\phi$ (relating the surface of the Earth to spatial infinity), whereas the experiments in question are finite-size laboratory experiments that make no explicit reference to spatial infinity.  

Within the general relativity community it is well-known that effects depending on the absolute gravitational potential $\phi$ should be treated with extreme care and discretion~\cite{Hell,Wald,wormholes}. Finite-size laboratories can only measure differences in gravitational potentials~\cite{observability}, so any effect that depends on the absolute gravitational potential $\phi$ is implicitly using an infinite size laboratory. One should be careful to verify just how (or whether) one is obtaining information from spatial infinity, and that one is including all contributions  to the absolute gravitational potential.
Unfortunately the calculations presented by the authors of~\cite{Morishima:2018a,Morishima:2018b,Morishima:2018c} fail both of these tests.

\section{Absolute gravitational potential}

The authors of~\cite{Morishima:2018a,Morishima:2018b,Morishima:2018c} note that the absolute gravitational potential at the surface of the Earth due to the presence of the Earth is $\phi_{Earth}/c^2 \approx 7\times 10^{-10}$. Unfortunately this is only part of the story: The Earth is deep inside the gravitational field of the Sun, and the 
absolute gravitational potential at the surface of the Earth due to the presence of the Sun is some 15 times larger $\phi_{Sun}/c^2 \approx 1\times 10^{-8}$. Even worse: The Earth (in fact the entire Solar system) is deep inside the gravitational field of the Galaxy, and the 
absolute gravitational potential at the surface of the Earth due to the presence of the Galaxy is some 2000 times larger than that obtained by just considering the Earth in isolation $\phi_{Galaxy}/c^2 \approx 1.4\times 10^{-6}$. In short, the 
absolute gravitational potential at the surface of the Earth is at least 2000 times greater than that estimated by the authors of~\cite{Morishima:2018a,Morishima:2018b,Morishima:2018c}. This vitiates their numerical estimates. 

Indeed the fact that it is the Galaxy that dominates the absolute gravitational potential at the surface of the Earth is a central reason that the general relativity community is extremely leery of any claims that local physics might depend on absolute gravitational potentials --- indeed it is a central tenet of general relativity that local physics should depend only on local gravitational effects. (This might possibly mean potential \emph{differences} across  the height of a finite-size laboratory, but in the context of general relativity is more likely to involve the spacetime curvature --- the Riemann tensor).

\section{Gravitational potential differences}

Using Newtonian gravity, specifically $\phi \sim - Gm/r$, the gravitational potential differences across a laboratory can be estimated as 
\begin{equation}
\Delta\phi \approx {d\phi\over dh} \; \Delta h \approx \phi  \;\;  {\hbox{(size of laboratory)}\over \hbox{(radius of Earth)}}.
\end{equation}
(Note that in \emph{potential gradients} the effect of the Sun and the Galaxy can safely be neglected, but not in the gravitational \emph{potentials}.) 
So even in Newtonian gravity, once one takes account of the local physics of a finite-size laboratory, there is an extra suppression factor of \hbox{(size of laboratory)}/\hbox{(radius of Earth)}. 
This is already enough to suppress the effects claimed in~\cite{Morishima:2018a,Morishima:2018b,Morishima:2018c} to undetectability. The situation is if anything worse within the context of general relativity.

\section{Einstein equivalence principle}

In general relativity the Einstein equivalence principle among other things implies the local flatness of spacetime and the existence of Riemann normal coordinates such that
\begin{equation}
g_{ab} = \eta_{ab} +  {1\over3}\;R_{acbd} \; x^c x^d + O( [x]^3).
\end{equation}
That is, you can always choose coordinates to locally make gravity ``go away'' up to a quadratic in coordinates term involving the Riemann tensor. This allows one to estimate the (fractional) gravitational effects due to spacetime curvature (the corrections to special relativity physics) as the dimensionless ratio
\begin{equation}
\hbox{(Riemann tensor)} \; \hbox{(size of laboratory)}^2.
\end{equation}
But the Riemann tensor is roughly the second derivative of the potential; so with $\phi\sim -Gm/r$ one has Riemann $\sim Gm/r^3 \sim \phi/r^2$. That is
\begin{equation}
\hbox{(Riemann tensor)}  \sim {\phi\over \hbox{(radius of Earth)}^2}
\end{equation}
and so in general relativity the fractional corrections to special relativity physics are
\begin{equation}
\hbox{(Riemann tensor)} \; \hbox{(size of laboratory)}^2 \approx \phi \; 
\left[ \hbox{(size of laboratory)}\over \hbox{(radius of Earth)}\right]^2.
\end{equation}
One is now, apart from the smallness of $\phi$,  fighting a quadratic suppression factor 
$\left[ \hbox{(size of laboratory)}/ \hbox{(radius of Earth)}\right]^2$.
This is an overwhelmingly small suppression factor that implies that gravitational contributions to laboratory based particle physics experiments are neglibly small. 

\section{Discussion}

While it is often mooted that gravity might have detectable laboratory-based particle physics effects~\cite{Morishima:2018a,Morishima:2018b,Morishima:2018c,Gharibyan:2014,Gharibyan:2012}, calculations are notoriously subtle and tricky.
Insofar as one obtains results depending on the absolute gravitational potential, then one is not doing general relativity, (and the gravitational effect of the Galaxy dominates over the Sun which in turn dominates over the Earth --- this is not a healthy state of affairs). In contrast, insofar as one working with standard general relativity, the Einstein equivalence principle implies the presence of overwhelmingly small suppression factors dependent on the size of the laboratory. 

\section*{Acknowledgments}
MV acknowledges financial support via the Marsden Fund administered by the Royal Society of New Zealand.


\end{document}